\documentclass[preprint,aps,showpacs]{revtex4}

\usepackage{amsmath,graphicx}

\begin{document}

\leftline{\hspace{5.8in} HD-THEP-04-7}

\title{\Large Nearly Minimal Magnetogenesis}

\author{\Large Tomislav Prokopec$^*$}
\author{Ewald Puchwein}
    \email[]{T.Prokopec@thphys.uni-heidelberg.de,
             Ewald.Puchwein@urz.uni-heidelberg.de}

\affiliation{Institut f\"ur Theoretische Physik, Heidelberg
             University, Philosophenweg 16, D-69120 Heidelberg, Germany\\}

\begin{abstract}

\bigskip

  We propose a new mechanism for magnetic field generation from inflation,
by which strong magnetic fields can be generated on cosmological scales.
These fields may be observable by cosmic microwave background radition 
measurements, and may have a dynamical impact on structure formation. The 
mechanism is based on the observation that a light nearly minimally coupled
charged scalar may be responsible for the creation of a negative photon 
mass-squared (provided the scalar field coupling to the curvature scalar is 
small but negative), which in turn results in abundant photon production
-- and thus in growing magnetic fields -- during inflation.

\end{abstract}

\pacs{98.80.Cq, 04.62.+v}

\vskip 0.4in

\maketitle

%
%

\section{Cosmological magnetic fields}
\label{Cosmological magnetic fields}

 There is a growing evidence that magnetic fields permeate intergalactic 
medium~\cite{Widrow:2002,Kronberg:1993,CarilliTaylor:2001,VogtEnsslin:2003}.
The galactic $\alpha-\Omega$ dynamo mechanism~\cite{Parker:1971} 
represents the standard explanation,
according to which small seed fields in protogalaxies are magnified
to a microgauss strength correlated on kiloparsec scales observed today
in many galaxies. Yet the dynamo does poorly as regards to explaning   
the observed cluster fields, which are typically correlated over
several kiloparsecs and can reach $10$ microgauss strength. 
On larger scales the field strength drops quite 
dramatically, and it is characterized by a negative 
spectral index between $-1.6$ and $- 2$~\cite{VogtEnsslin:2003},
consistent with the Kolmogorov spectrum of turbulence.

The seed field that fuels the dynamo, which amplifies 
the galactic (and possibly cluster) fields, has either 
primordial origin, or it was created by the Biermann battery mechanism,
operative at the time of structure and galaxy formation,
when the Universe was about one billion years 
old~\cite{PudritzSilk:1989,KulsrudCenOstrikerRyu:1996,DaviesWidrow:2000}.
The Biermann battery mechanism is operative in the presence of 
oblique shocks, which generate nonideal fluid that violates  
proportionality between the gradients of pressure and energy density.
Such a fluid can produce the vorticity required
for generation of seed fields, whose strength is typically
of the order $10^{-20}$~gauss.
While in the protogalactic environments these seed fields may
be sufficiently strong to fuel the galactic dynamo mechanism, 
whether they can be used to explain the observed cluster fields
is a highly controversial question. Moreover, these fields may be
insufficiently strong to explain the galactic
fields observed in a couple of galaxies 
at higher redshifts~\cite{Kronberg:1993,Widrow:2002}.

 The quest for the fields correlated on much larger scales has so far been 
unsuccessful. A notable exception is the result 
of~\cite{AnchordoquiGoldberg:2001}, where a local field of about
$10^{-8}$ gauss correlated on  megaparsec scale is quoted. 
If established, the existence of extragalactic fields would point
at a primordial origin. 
Perhaps the most promising method for detecting 
primordial fields correlated over supragalactic scales
is through their impact on the cosmic microwave background radiation 
(CMBR)~\cite{BarrowFerreiraSilk:1997,DurrerFerreiraKahniashvili:1999},  
and possibly through their dynamical influence
on large scale structure and galaxy formation.
The current upper bounds from the CMBR are of the order 
$10^{-9}$~gauss, while the bounds from structure formation are somewhat
weaker. Fields stronger than about $10^{-5}$ gauss 
at the time of galaxy formation
(which correspond to about $10^{-9}$~gauss comoving field strength)
are dynamically relevant.

  In this Letter we propose a novel mechanism for generation of large
scale magnetic fields during cosmological inflation.
Our mechanism is based on a model with a rather minimal set of assumptions:
We study the dynamics of (one-loop) quantum  scalar
electrodynamics~\cite{ProkopecPuchwein:2003,ProkopecTornkvistWoodard:2002AoP,
ProkopecTornkvistWoodard:2002prl,ProkopecWoodard:2003ajp,
ProkopecWoodard:2003photons} in cosmological space-times, 
which can be naturally embedded into 
the Standard Model (where the role of the scalars is played by
the charged components of the Higgs field), or into supersymmetric
extensions of the Standard Model (the charged scalars are the 
sypersymmetric partners of the Standard Model fermions).


\section{Scenario} 
\label{Scenario}

We assume a standard model of the Universe,
in which a period of inflation is followed by radiation and matter eras. 
The background space-time during inflation can be accurately approximated 
by the de Sitter metric, $g_{\mu\nu} = a^2\eta_{\mu\nu}$
($\eta_{\mu\nu} = {\rm diag}[-1,1,1,1]$)
with the scale factor $a=-1/(H\eta)$ (for $\eta< - \eta_H \equiv - 1/H$), where
$\eta$ denotes conformal time, and 
$H$ the inflationary Hubble parameter. 
After inflation the Universe undergoes a sudden transition to radiation
domination, in which $a=H\eta$ (for $\eta>\eta_H$). 
When inflation lasts a sufficiently long time, 
such that a de Sitter invariant solution is
established, the dynamics of photons that couple to {\it light} scalar
particles, which in turn couple {\it nearly minimally} to gravity, can
be described by the Proca Lagrangian \cite{ProkopecPuchwein:2003}
\vskip -0.15in
\begin{equation}
 {\cal L}_{\rm Proca} \!=\! - \frac 14 \eta^{\mu\rho}\eta^{\nu\sigma}
                                         F_{\mu\nu}F_{\rho\sigma}
                   \!- \frac 12 a^2 m_\gamma^2 \eta^{\mu\nu}
                                  \! A_{\mu} A_{\nu}
                   \!+ O({\tt s}^0),\!\!
\label{Proca:L}
\end{equation}
\vskip -0.02in
\noindent
where 
$F_{\mu\nu}=\partial_\mu A_\nu-\partial_\nu A_\mu$. 
The photon mass is given by
\vskip -0.15in
\begin{equation}
    m_{\gamma}^2=\frac{\alpha H^2}{\pi {\tt s}}
\,,\quad 
    \quad
         {\tt s}=\frac{m_\Phi^2+ \xi {\cal R}}{3H^2},
\end{equation}
\vskip -0.05in
\noindent
where $\alpha = e^2/4\pi$ is the fine structure constant,
${\cal R}=12H^2$ is the curvature scalar of de Sitter space, $m_\Phi$ is
the scalar mass and $\xi$ describes the coupling of the scalar
field to ${\cal R}$. When $0<m_\Phi^2\ll H^2$ and $-1/12\ll \xi<0$, 
both $\tt s$ and $m_{\gamma}^2$ can be negative. We emphasise that 
there is nothing patological about this limit. Indeed, 
the de Sitter invariant (Feynman) scalar propagator
$i\Delta_F(x,x^\prime) \equiv G_F(y)$ in $D=4$
reads~\cite{ProkopecPuchwein:2003}
\begin{equation}
 G_F(y)  =   \frac{H^2}{4\pi^2}\bigg\{
                              \frac{1}{y}
                            - \frac 12 \ln(y)
                            + \frac 1{2{\tt s}}
                            - 1 + \ln(2)
                            + O({\tt s})
                    \bigg\}
\,,
\label{iDelta:4dim:massive}
\end{equation}
where $y(x;x') \equiv a a' H^2 {\Delta x}^2(x;x')$
denotes the de Sitter invariant length, and 
${\Delta x}^2(x;x') = -(|\eta - \eta^\prime| - i\epsilon)^2 
                   + \|\vec x - \vec x\,'\|^2$, $\epsilon\rightarrow 0^+$,
such that a negative small ${\tt s}$ introduces negative, but finite,
correlations, which are at the heart of the mechanism discussed in this Letter.

 The Lagrangean~(\ref{Proca:L}) was derived
in~\cite{ProkopecPuchwein:2003} by
expanding in powers of $|{\tt s}|\ll 1$ 
the one-loop (order $\alpha$) nonlocal effective action written in 
the Schwinger-Keldysh formalism. 
In~\cite{ProkopecPuchwein:2003} we proved that, when written
in the generalized Lorentz gauge $\partial_\mu\big(a^2\eta^{\mu\nu} A_\nu\big) = 0$
at the order $O(\alpha,{\tt s}^{-1})$,
the effective action reduces to the Proca theory~(\ref{Proca:L}).

Due to spatial homogeneity, the general solution for the photon field
of the Proca equation can be conveniently written as a superposition of
spatial plane waves 
$A_{\mu}(x)=\varepsilon_{\mu}(\vec{k},\eta){\rm e}^{i\vec{k}\cdot\vec{x}}$, where
$\varepsilon_\mu=(\varepsilon_0,\vec{\varepsilon})$.
Then $\varepsilon_{0}$ is nondynamical and just traces the spatial
components. The spatial componets correpond to the three physical degrees of 
freedom, and can be decomposed into longitudinal
$\vec{\varepsilon}_{{\tt L},\vec{k}}
  \equiv \vec{k}\, (\vec{k}\cdot \vec{\varepsilon})/\vec{k}^2$ 
and transverse part
$\vec{\varepsilon}_{{\tt T},\vec{k}}\equiv\vec{\varepsilon}
        -\vec{\varepsilon}_{{\tt L},\vec{k}}$. 
From Eq.~(\ref{Proca:L}), one finds the
following equation of motion for the transverse polarization
$\vec{\varepsilon}_{{\tt T},\vec{k}}$ during inflation
\begin{equation}
(\partial_\eta^2+k^2+m_{\gamma}^2a^2)\,\vec{\varepsilon}_{{\tt T},\vec{k}}(\eta)
   =0
\,,
\end{equation}
where $k\equiv\|\vec{k}\|$.
Demanding that the solution of this equation corresponds 
to a (circularly polarized) vacuum state for
$\eta\rightarrow -\infty$, one finds
$\vec{\varepsilon}_{{\tt T},\vec{k}}(\eta)=
A_{\vec{k}}(\eta)(\vec{\epsilon}_{\vec{k}}^{\,1}
                          + i\vec{\epsilon}_{\vec{k}}^{\,2})/\sqrt{2}$,
where  $\vec{\epsilon}_{\vec{k}}^{\,\cal T}$ (${\cal T}=1,2$) are the
transverse polarization vectors and
\begin{equation}
    A_{\vec{k}} = \frac{1}{2}(-\pi\eta)^{\frac 12}\mathrm{H}_\nu^{(1)}(-k\eta)
\,,\qquad
    \nu = \sqrt{\frac{1}{4} -\frac{m_\gamma^2}{H^2}}
\,,
\label{eq - solution, de Sitter}
\end{equation}
where $\mathrm{H}_{\,\nu}^{(1)}$ is the Hankel function.
The second solution is simply
$\vec{\varepsilon}_{{\tt T},\vec{k}}^{\,*}(\eta)$, such that 
the Wronskian $W[\vec{\varepsilon}_{{\tt T},\vec{k}},
                  \vec{\varepsilon}_{{\tt T},\vec{k}}^{\,*}] = i$.


\section{Vanishing conductivity}
\label{Vanishing conductivity}

When conducitivity in radiation era is negligibly small,
a smooth matching of the inflationary epoch 
solution~(\ref{eq - solution, de Sitter}) to the radiation era
modes 
\begin{equation}
    A_{\vec{k}}^\pm(\eta)=\frac {1}{\sqrt{2k}}\,{\rm e}^{\mp ik\eta}
\label{eq - solution, radiation}
\end{equation}
leads to the radiation era 
solution~({\it cf.} Ref.~\cite{DavisDimopoulosProkopecTornkvist:2000}).
By matching Eq.~(\ref{eq - solution, de Sitter}) to a linear
combination of these solutions, one finds
\begin{align}
    {\cal A}_0
   &=\alpha_k A_{\vec{k}}^+|_{\eta
   =\eta_H}+\beta^*_kA_{\vec{k}}^-|_{\eta=\eta_H}
\,,
\\
    {\cal A}^\prime_0
      & =\alpha_k(\partial_\eta A_{\vec{k}}^+)|_{\eta
        =\eta_H}+\beta^*_k(\partial_\eta A_{\vec{k}}^-)|_{\eta=\eta_H}
\,,
\end{align}
where ${\cal A}_0 \equiv A_{\vec{k}}|_{\eta=-\eta_H}$ and 
${\cal A}^\prime_0 \equiv (\partial_\eta A_{\vec{k}})|_{\eta=-\eta_H}$.
Solving for $\alpha_k$, $\beta^*_k$ gives
\begin{align}
    \alpha_k = \frac{{\cal A}_0 +({i}/{k}){\cal A}^\prime_0}
            {2A_{\vec{k}}^+|_{\eta=\eta_H}}
\,,\qquad
    \beta^*_k= \frac{{\cal A}_0 - ({i}/{k}){\cal A}^\prime_0}
                             {2A_{\vec{k}}^-|_{\eta=\eta_H}}
\,,
\label{eq - alpha-beta-k}
\end{align}
such that the gauge field evolves in radiation era as 
\begin{equation}
    A_{\vec{k}}(\eta) = {\cal A}_0 \cos k(\eta-\eta_H)
                      + \frac{{\cal A}_0^\prime}{k}\sin k(\eta-\eta_H)
\quad(\eta>\eta_H)
\label{A:radiation}
\,.
\end{equation}
For modes that are superhorizon at the end of inflation ($k\ll H$)
one can use the small-argument expansion of the Hankel function 
$H_\nu^{(1)}(k/H)$ to get (for $\nu> 0$)
%
%
\begin{eqnarray}
  {\cal A}_0 &=& - i\frac{\Gamma(\nu)}{\sqrt{2\pi}}(2H)^{\nu-\frac 12} k^{-\nu}
             + O(k^\nu,k^{-\nu+1})
\label{calA0}
\\
   {\cal A}_0^\prime &=&  \Big(\nu -\frac 12\Big) H {\cal A}_0 
\label{calA0'}
\,.
\end{eqnarray}
This result is used below to calculate the magnetic field spectrum
in radiation era.
%
Since matter era is a conformal space-time, the fields (as well as their spectra)
in matter era are inherited from the radiation era fields~(\ref{A:radiation}).


\section{Large conductivity}
\label{Large conductivity}

An opposite extreme is a sudden increase of conductivity in radiation era,
which occurs for example in the case of a rapid thermalisation after inflation,
resulting in a large (thermal) conductivity. 
To leading logarithm (squared) in the coupling constant,
the evolution of the photon field
is then governed by the B\"odeker-Langevin equation~\cite{Bodeker:1998hm}
\begin{equation}
     (a\sigma\partial_\eta+k^2)
       \varepsilon_{\vec{k}}^{\cal T}(\eta)
         = a^3\xi^{\,\cal T}(\vec{k},\eta)
\,,
    \label{eq - eom large sigma}
\end{equation}
where $\varepsilon_{\vec{k}}^{\cal T}$ is defined by 
$\vec \varepsilon_{{\tt T},\vec k}=
   \sum_{\cal T}\varepsilon_{\vec{k}}^{\cal T}\vec 
                       \epsilon_{\vec{k}}^{\,\cal T}$.
The stochastic force $\xi^{\,\cal T}$ satisfies
the following Markowian fluctuation-dissipation relation
\begin{equation}
   \langle\xi^{\cal T}(\vec{k},\eta)\xi^{\cal T\,'*}(\vec{k}',\eta')\rangle
     \mspace{-5mu} = \mspace{-3mu}
         \frac{2\sigma T}{a^4}
    (2\pi)^3\delta^{\cal TT\,'}\mspace{-5mu}
              \delta(\eta-\eta')\delta^{(3)}(\vec{k}-\vec{k}')
\,,
    \label{eq - fluctuation-dissipation}
\end{equation}
where 
$T$ denotes the equivalent temperature (average energy of the
plasma excitations). The factor $a^3$, which multiplies
$\xi^{\,\cal T}$ in Eq.~(\ref{eq - eom large sigma}), can be
inferred from the fact that the transverse components of the
current density are given by $j^{\,\cal T}=\sigma E^{\,\cal T}+\xi^{\,\cal T}$,
and from the transformation properties of the
conductivity and the electric field. 
Eq.~(\ref{eq - eom large sigma}) can be easily integrated,
\begin{equation}
    \varepsilon_{\vec{k}}^{\cal T}(\eta)
    =\mspace{-5mu} \int_{\eta_H}^\eta\mspace{-10mu} 
{\rm exp}\Big({-\mspace{-5mu}
         \int_{\eta'}^{\eta}d\eta''{k^2}/({a\sigma})
         }\Big)
    \frac{a^2\xi^{\cal T}}{\sigma}d\eta'
+ {\rm exp}\Big({-\mspace{-1mu}
         \int_{\eta_H}^{\eta}\mspace{-5mu}\frac{k^2}{a\sigma}d\eta'}\Big)
             \varepsilon^{\cal T}_{\vec{k}}(\eta_H)
\,.
\end{equation}
Using this and~(\ref{eq - fluctuation-dissipation}) one can
derive the equal time correlator
\begin{eqnarray}
\langle
 \varepsilon_{\vec{k}}^{\cal T}(\eta)\varepsilon_{\vec{k}'}^{\cal T\,'*}(\eta)
\rangle
&=& 
 \int_{\eta_H}^\eta\! d\eta'\,
                 {\rm exp}\Big(
                      -2\int_{\eta'}^{\eta}d\eta''\frac{k^2}{a\sigma}
                          \Big)
    \frac{2T}{\sigma}
 (2\pi)^3\delta^{\cal TT\,'}\delta^{(3)}(\vec{k}-\vec{k}')\mspace{-10mu}
\nonumber\\
\!\!   &+& {\rm exp}\Big(
                       {-\int_{\eta_H}^{\eta}d\eta'\frac{k^2+k'^2}{a\sigma}}
                  \Big)
    \varepsilon_{\vec{k}}^{\cal T}(\eta_H)\varepsilon_{\vec{k}'}^{\cal
    T\,'*}(\eta_H).
\end{eqnarray}
Assuming $\sigma,T\propto 1/a$ (which is justified in and close to
thermal equilibrium), one can perform the integrations, to obtain
\begin{eqnarray}
    \langle\varepsilon_{\vec{k}}^{\cal T}(\eta)\varepsilon_{\vec{k}'}^{\cal
    T\,'*}(\eta)\rangle &=& \frac{aT}{k^2}
         \Big[1
             - {\rm exp}\Big({-\frac{2k^2}{a\sigma}\Delta\eta}\Big)
          \Big]
               (2\pi)^3\delta^{\cal TT\,'}\delta^{(3)}(\vec{k}\!-\!\vec{k}')
\nonumber\\
    &+&\, {\rm exp}\Big(
                     {-\frac{k^2+k'^2}{a\sigma}\Delta\eta}
                \Big) 
                \varepsilon_{\vec{k}}^{\cal T}(\eta_H)
                   \varepsilon_{\vec{k}'}^{\cal T\,'*}(\eta_H)
\,,\quad
\label{correlator:radiation}
\end{eqnarray}
where $\Delta\eta=\eta-\eta_H$, such that the spectrum is neatly split into
the thermal and primoridial contribution.


\section{Magnetic field spectrum} 
\label{Magnetic field spectrum}

When suitably averaged over a physical scale $\ell_{\rm ph} = a\ell$,
the magnetic field operator can be defined as
\vskip -0.2in
\begin{equation}
  \hat{\vec B}_\ell(\eta,\vec x) \equiv \int d^3 y \,
           \mathtt{W}(\vec x-\vec y,\ell) \hat{\vec B}(\eta,\vec y)
\,,
\label{averaged B operator}
\end{equation}
where $\mathtt{W}$ denotes a coordinate space window function,
which, {\it e.g.}, can be chosen as
$\mathtt{W}= [2\pi\ell^2]^{-3/2}\exp 
   \big(-||\vec x -\vec y\,||^2/(2\ell^2)\big)$,
and 
\begin{equation}
 \hat{\vec B}(\eta,\vec y) = \frac{1}{a^2}\int\! \frac{d^3k}{(2\pi)^3}
   i\vec k\times \sum_{\cal T}
     \Big[
      A_{\vec{k}}(\eta)\,\vec{\epsilon}_{\vec{k}}^{\,\,\cal T}
                 \,{\rm e}^{i\vec{k}\cdot\vec{y}}\,\hat a^{\cal T}_{\vec k}
          - {\rm h.c.}
     \Big]
.\!
\label{B:radiation}
\end{equation}
The creation and annihilation operators $\hat a_{\vec k}^{\cal T\dagger}$
and $\hat a_{\vec k}^{\cal T}$ obey the commutation relation,
$[\hat a_{\vec k}^{\cal T},\hat a_{\vec k'}^{\cal T'\dagger}] 
      = (2\pi)^3\delta^{\cal T T'}\delta^{(3)}(\vec k-\vec k')$, and 
$\hat a_{\vec k}^{\cal T}|0\rangle = 0$. When~(\ref{averaged B operator})
is squared and averaged over the vacuum state, one arrives at
\begin{eqnarray}
 \langle \vec B_\ell^2(\eta,\vec x)\rangle 
          &\equiv&  \int \frac{dk}{k}|{\cal W}|^2
 {\cal P}_B
,\quad
   {\cal P}_B = \frac{1}{a^4}\frac{k^5}{\pi^2}|A_{\vec k}(\eta)|^2
,\qquad
\label{spectrum:definition}
\label{spectrum:definition2}
\end{eqnarray}
where ${\cal P}_B={\cal P}_B(\eta,k)$
%
%
defines the {\it magnetic field spectrum},
and  ${\cal W}(\ell,k) 
      = \int d^3 z\, \mathtt{W}(\vec z,\eta)\exp(i\vec k\cdot\vec z)$ 
is the momentum space window function
(${\cal W} = \exp(-k^2\ell^2/2)$ for the
 coordinate space window function 
$\mathtt{W}$ mentioned above). 

\begin{figure}[h]
\begin{center}
  \hskip -0.2in
  \includegraphics[scale=1]{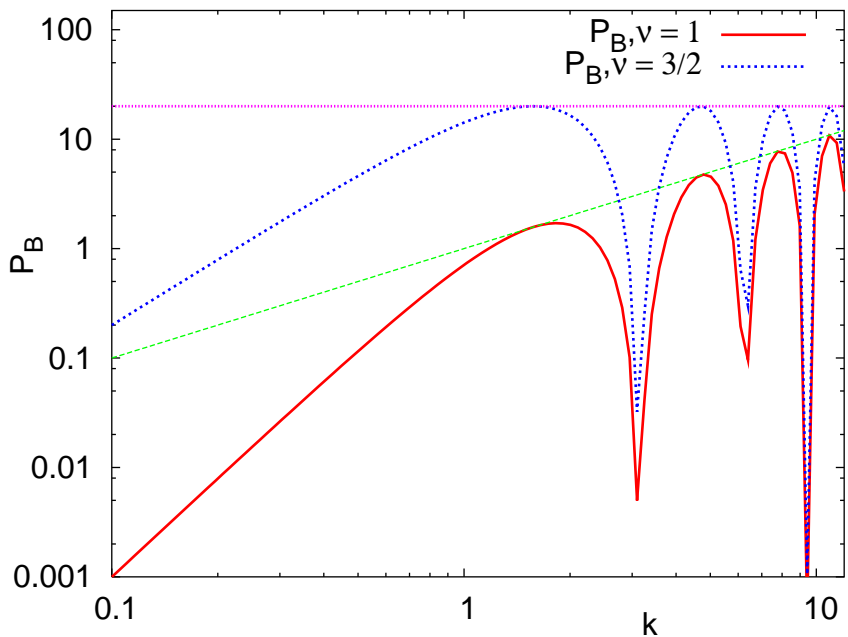}
\end{center}
  \vskip -0.3in
\caption{Magnetic field spectrum in radiation era~(\ref{spectrum:radiation}) 
for vanishing conductivity on a log-log scale for $\nu =1$ and  $\nu =3/2$.
(Normalization is arbitrary.)}
    \label{fig1:spectrum:sigma=0}
\end{figure}

When conductivity is vanishingly small, the spectrum in radiation era
can be calculated by inserting Eqs.~(\ref{A:radiation}--\ref{calA0'})
into~(\ref{spectrum:definition2}). The result is (when $a\gg 1$)
\begin{eqnarray}
 {\cal P}_B \simeq \frac{1}{a^4}\frac{\Gamma^2(\nu)}{(2\pi)^3} 
                 \frac{(2H)^{2\nu+1}}{k^{2\nu-3}} \big(\nu-\frac 12 \big)^2
   \sin^2\{ k(\eta-\eta_H)\} 
\,,\quad
\label{spectrum:radiation}
\end{eqnarray}
%
such that on superhorizon scales $k\eta \ll 1$
the spectrum~(\ref{spectrum:radiation}) ${\cal P}_B \propto k^{5-2\nu}$, 
and on subhorizon scales the spectrum is oscillatory, 
with the envelope scaling as ${\cal P}_B \propto k^{3-2\nu}$. 
This spectrum is shown in figure~\ref{fig1:spectrum:sigma=0} on a log-log plot. 
Note that  scale invariance is reached for
$\nu_{\rm flat} = 5/2$ (superhorizon scales)
$\nu_{\rm flat} = 3/2$ (subhorizon scales).

 In the limit of a large conductivity $\sigma$, the magnetic field spectrum 
can be split into the contribution from thermal excitations, and 
the primordial contribution. A new scale arises 
in Eq.~(\ref{correlator:radiation}),
associated with the momentum
%
\begin{equation}
k_{\sigma} = \sqrt{{\sigma H}/{2}}
\,,
\label{conductivity scale}
\end{equation}
which we refer to as the {\it conductivity scale}.
The thermal spectrum can be inferred from~(\ref{correlator:radiation}):
In the ultraviolet $P_B^{\rm th}\propto k^3$  ($k\gg k_\sigma$),
while in the infrared  $P_B^{\rm th}\propto k^{5}$ ($k\ll k_\sigma$), 
implying that thermal excitations give rise to fields of negligible strength
on cosmological scales 
(see~\cite{DavisDimopoulosProkopecTornkvist:2000}). 

\begin{figure}[h]
\begin{center}
  \hskip -0.2in
  \includegraphics[scale=1]{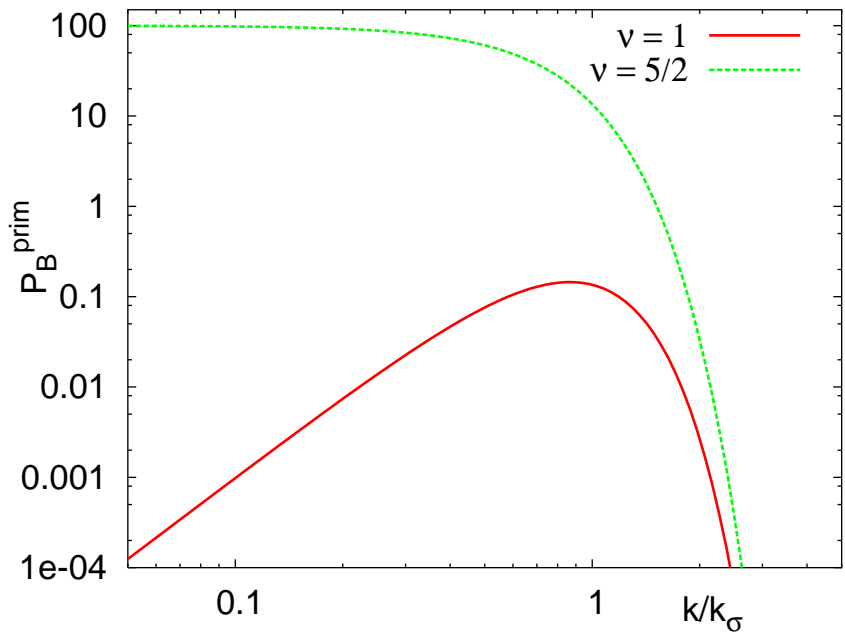}
\end{center}
  \vskip -0.3in
\caption{Magnetic field spectrum in 
radiation~(\ref{spectrum:large sigma}) when a large conductivity 
sets in rapidly after inflation
on a log-log scale for $\nu =1$ and  $\nu =5/2$.
(Normalization is arbitrary.)}
    \label{fig2:spectrum:sigma}
  \vskip -.15in
\end{figure}

 The primordial contribution to the spectrum can be obtained from 
Eqs.~(\ref{calA0}), (\ref{correlator:radiation}) 
and~(\ref{spectrum:definition2}),
\vskip -0.15in
\begin{eqnarray}
 {\cal P}_B^{\rm prim} \simeq \frac{1}{a^4}\frac{\Gamma^2(\nu)}{2\pi^3} 
                 \frac{(2H)^{2\nu-1}}{k^{2\nu-5}}
                   {\exp}\Big(\!-\frac{2k^2}{H\sigma}(1\!-\!a^{-1})\Big)
,\;
\label{spectrum:large sigma}
\end{eqnarray}
\vskip -0.05in
\noindent
such that the spectrum is exponentially cut-off when 
$k>k_\sigma$. The flat spectrum is reached for $\nu_{\rm flat} = 5/2$
for $k<k_\sigma$.
This spectrum is shown in figure~\ref{fig2:spectrum:sigma} on a log-log plot. 
Since typically in radiation era $\sigma \sim T\gg H/a$ ($a\gg 1$), 
 $k_{\sigma,\rm ph}\equiv k_\sigma/a \gg H(t) = H/a^2$, 
implying that a rapid growth in conductivity 
after inflation freezes out large scales magnetic fields 
and destroys the small scales fields
(electric field spectrum is completely destroyed), such that no oscillations
are present on subhorizon scales. Thus, the absence or presence of 
subhorizon oscillations, and observation of a conductivity scale, 
can be testing grounds for the conductivity history during radiation era.


\section{Discussion}
\label{Discussion}

The spectrum~(\ref{spectrum:radiation}) can be thought of as a function of 
the scalar coupling to gravity $\xi$ and its mass $m_\Phi$, which are 
fundamental parameters characterizing a scalar field. To illustrate this point
more precisely, we show the spectral index $n = 3 - 2\nu$ 
characterizing the spectrum envelope of 
  ${\cal P}_B\propto k^n$ in Eq.~(\ref{spectrum:radiation}) 
on subhorizon scales, $k/a > H(t)$
(or $B_\ell \equiv \langle\vec B_\ell^2\rangle^{1/2}
                    \propto \ell^{-n/2}$ ($n\neq 0$),
  $B_\ell \propto -\ln(\ell)$ ($n = 0$) in Eq.~(\ref{spectrum:definition})).
Between $\xi = \xi_{\rm crit} \equiv - m_\Phi^2/(12H^2)$  and  
$\xi = \xi_3 \equiv (\alpha/\pi) - m_\Phi^2/(12H^2)$,
$n=3$, equal to that of thermal spectrum.
Above $\xi > \xi_3$, $n$ drops, reaching an asymptotic value 
$n\rightarrow n_{\infty} = 2$ when $\xi \rightarrow \infty$. 
The spectral index on superhorizon scales is obtained simply by replacing
$n + 2 \rightarrow n = 5-2\nu$.
When compared with the vacuum spectrum,  ${\cal P}_{B}^{\rm vac} \propto k^4$
($k<H$),
a spectrum with $n \in (2,3)$ exhibits an enhancement
of magnetic fields on subhorizon scales, which is
due to a conversion of the electric energy (which is enhanced during inflation)
into the magnetic energy during radiation era. The conversion is efficient 
provided radiation era is characterised by a low conductivity. 
This mechanism was used in
Refs.~\cite{ProkopecPuchwein:2003,ProkopecTornkvistWoodard:2002AoP,
ProkopecTornkvistWoodard:2002prl,ProkopecWoodard:2003ajp,
ProkopecWoodard:2003photons,DavisDimopoulosProkopecTornkvist:2000,
DimopoulosProkopecTornkvistDavis:2001} 
to argue that a spectrum $B_\ell \propto \ell^{-1}$ 
 can be obtained 
from inflation, which could be sufficient to seed the galactic dynamo
mechanism.

When $\xi\in(-\infty,\xi_{\rm crit})$ however, 
gauge fields exhibit instability and are enhanced during inflation,
such that the spectral index of subhorizon modes drops from 
$n=2$ when $\xi \rightarrow -\infty$, to flat spectrum ($n=0$), when 
\begin{equation}
    \xi_{\rm flat} = -\frac{\alpha}{8\pi}-\frac{m_\Phi^2}{12H^2}
\,,
\label{xi flat}
\end{equation}
to $n<0$, when $\xi \in (\xi_{\rm flat},\xi_{\rm crit})$.  
A negative spectral index $n \leq - 2$ would imply 
a growth in magnetic field energy during inflation, 
resulting in a  divergent magnetic (and electric) field energy,
as it can be inferred from Eqs.~(\ref{spectrum:radiation}), 
(\ref{spectrum:large sigma}) 
and~(\ref{spectrum:definition}),
with ${\cal W} = 1$. A proper study of this case would require
an inclusion of backreaction of the electromagnetic field on the background
space-time, which is beyond the scope of this Letter.
 A spectral index $n > -2$ on subhorizon scales 
(or $n > 0$ on superhorizon scales), 
results in an acceptable magnetic field spectrum
on cosmological scales, whose dynamical impact on CMBR and
large scale structure formation should be considered. 
\begin{figure}[h]
\begin{center}
  \hskip -0.05in
  \includegraphics[scale=1]{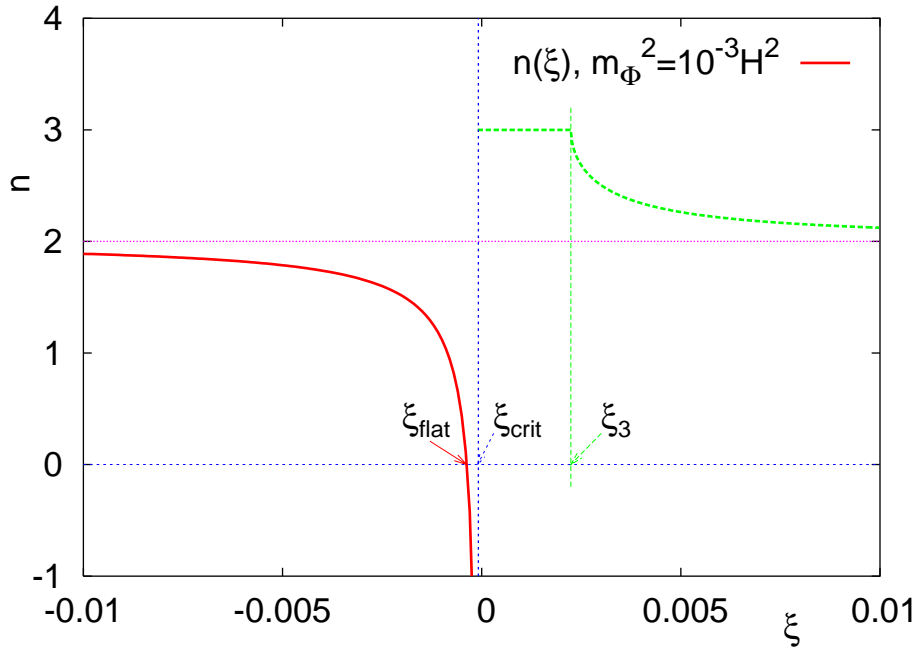}
\end{center}
  \vskip -0.25in
\caption{Spectral index $n$ for subhorizon modes in the low conductivity case
   ({\it cf.} Eq.~(\ref{spectrum:radiation})) as a function of the scalar coupling
    to gravity $\xi$ 
   (with $\alpha = 1/137$ and the scalar mass $m_\Phi^2 = 10^{-3}H^2$).
}
    \label{fig3:spectral index}
\end{figure}

 Spectrum normalization today can be determined from 
magnetic field strength at a comoving scale corresponding 
to the Hubble scale at the end of inflation ($k=H$).
At the comoving inflation scale $\ell_H\sim 3~{\rm m} \sim 10^{-16}$ parsec
(corresponding to $H \sim 10^{37}~{\rm Hz} \sim 10^{13}~{\rm GeV}$)
the field strength is about 
$B_H \sim 10^{-12}-10^{-11}$~gauss~\cite{Prokopec:2001}.
Therefore, if the spectrum is (almost) flat,
the magnetic field is potentially observable 
by the next generation of CMBR experiments (PLANCK satellite), as well
as of importance for the dynamics of large 
scale structures of the Universe. 

\bigskip

\acknowledgements{We would like to thank Richard P. Woodard for discussions.
E.P. would like to thank Michael G. Schmidt for guidance in his diploma thesis,
which inspired this work.}

%
%

\end{document}